\documentclass[10pt]{iopart}

\usepackage{epsfig,iopams}

\begin{document}

\title[Mortensen {\it et al.}, Small-core photonic crystal fibers ....]{Small-core photonic crystal fibers with weakly disordered air-hole claddings}

\author{Niels Asger Mortensen}
\address{Crystal Fibre A/S, Blokken 84, DK-3460 Birker\o d, Denmark}

\author{Martin D. Nielsen}
\address{Crystal Fibre A/S, Blokken 84, DK-3460 Birker\o d, Denmark\\COM, Technical University of Denmark, DK-2800 Kongens Lyngby, Denmark}

\author{Jacob Riis Folkenberg}
\address{Crystal Fibre A/S, Blokken 84, DK-3460 Birker\o d, Denmark}

\author{Kim P. Hansen}
\address{Crystal Fibre A/S, Blokken 84, DK-3460 Birker\o d, Denmark\\COM, Technical University of Denmark, DK-2800 Kongens Lyngby, Denmark}

\author{Jesper L{\ae}gsgaard}
\address{COM, Technical University of Denmark, DK-2800 Kongens Lyngby, Denmark}

\begin{abstract}
Motivated by recent experimental work by Folkenberg {\it et al.} we consider the effect of weak disorder in the air-hole lattice of small-core photonic crystal fibers. We find that the broken symmetry leads to higher-order modes which have generic intensity distributions resembling those found in standard fibers with elliptical cores. This explains why recently reported experimental higher-order mode profiles appear very different from those calculated numerically for ideal photonic crystal fibers with inversion and six-fold rotational symmetry. The splitting of the four higher-order modes into two groups fully correlates with the observation that these modes have different cut-offs. 
\end{abstract}

\submitto{\JOA}

\maketitle

The advances in fabrication techniques and structural control of both index-guiding as well as photonic band-gap photonic crystal fibers (PCFs) is believed to improve the agreement between experiments and theory based on ideal structures (for a recent review of the field we refer to Ref.~\cite{russell2003} and references therein). The recent investigation of micro-bending induced attenuation in PCFs~\cite{nielsen2003} is an example of the successful merging of experiments and simulations as well as recent experiments~\cite{folkenberg2003} on multi-mode cut-off in small-core PCFs and the good agreement with theoretical predictions~\cite{mortensen2003c,kuhlmey2002}. On the other hand, in the latter experiments the observed higher-order mode profiles were very different from those of an ideal structure despite the fact that the studied fibers appeared to have highly regular air-hole lattices. However, the relative strength of disorder may still be significant and in this Letter we numerically study the effect of such variations on both fields and propagation constants.

\begin{figure}[b!]
\begin{center}
\epsfig{file=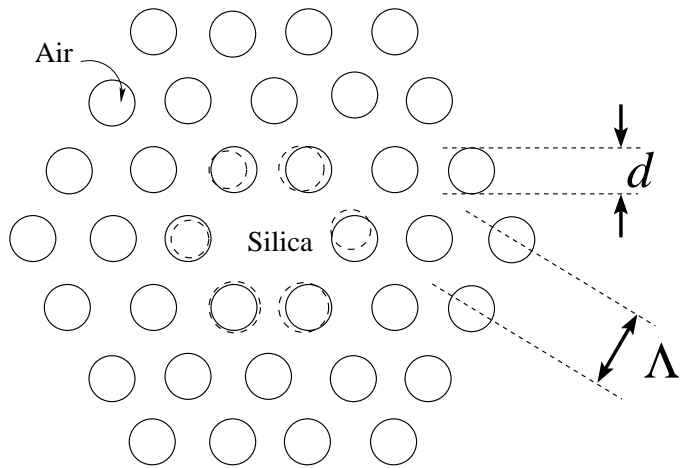, width=0.7\textwidth,clip}
\end{center}
\caption{Schematic of the cross-section of a photonic crystal fiber. The dashed circles indicate small random displacements and variations in diameter of the six inner air holes. }
\label{fig1}
\end{figure}

\begin{figure*}[t!]
\begin{center}
\epsfig{file=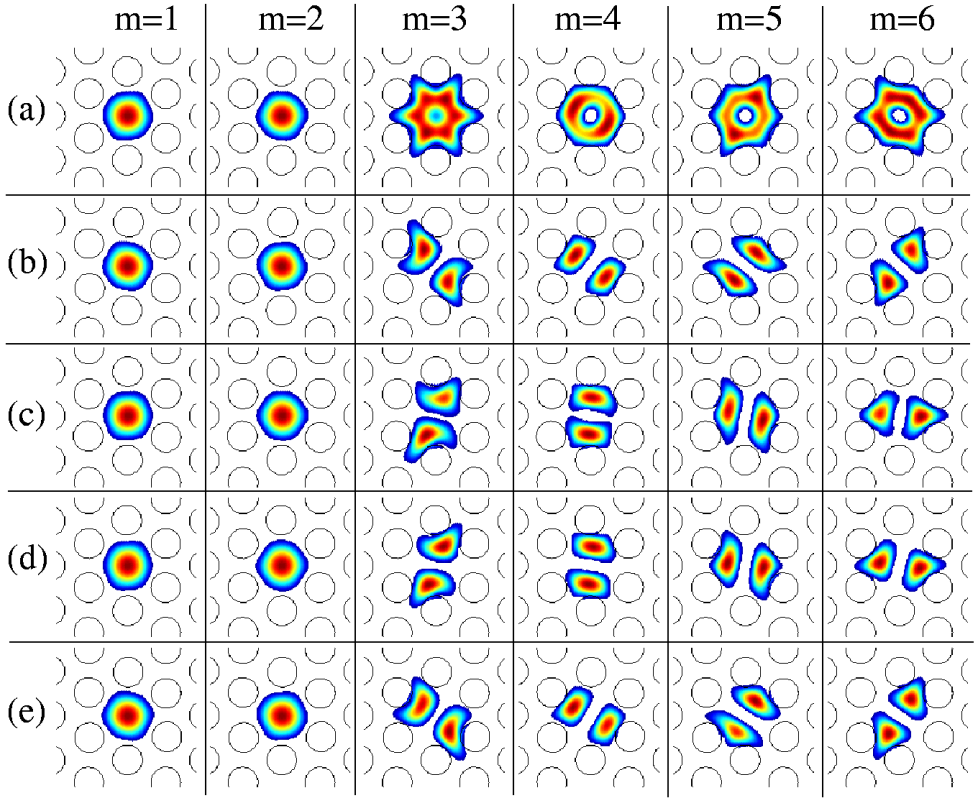, width=0.99\textwidth,clip}
\end{center}
\caption{Plot of the electrical field intensity. Row (a) shows results for the ideal structure and rows (b)-(e) are for different random disorder in the inner ring of the air-hole lattice.}
\label{fig2}
\end{figure*}

We consider the type of PCF first studied by Knight {\it et al.}~\cite{knight1996} for which the ideal structure has inversion and six-fold rotational symmetry, see Fig.~\ref{fig1}. In the first row of Fig.~\ref{fig2} we show the electrical field intensity $|{\boldsymbol E}_m^{(0)}|^2$ (the superscript emphasizes the ideal or ``unperturbed'' structure) of the first six ($m=1,2,3\ldots 6$) eigenmodes at $\lambda=780\,{\rm nm}$ of an ideal PCF with $\Lambda=1.4\,{\rm \mu m}$ and $d/\Lambda=0.67$ corresponding to one of the fibers studied in Ref.~\cite{folkenberg2003}. The results are based on a numerical fully-vectorial solution of Maxwell's equations in a plane-wave basis with periodic boundary conditions~\cite{johnson2001} and basis vectors which support the inversion and six-fold rotational symmetry of the dielectric structure. For the refractive index $n$ we use $n=1$ in air and for silica we use a Sellmeier expression with $n=1.45367$. The fundamental modes ($m=1$ and 2) have close-to-Gaussian profiles~\cite{mortensen2002a,mortensen2002b}.

Historically, their degeneracy has been widely debated (see Refs.~\cite{steel2001,koshiba2001} and references therein) but recently it was finally proved by Steel {\it et al.}~\cite{steel2001} how their degeneracy follows from group-theory because ${\boldsymbol E}_1^{(0)}$ and ${\boldsymbol E}_2^{(0)}$ do not support the full symmetry. Recently, the degeneracies of higher-order modes were addressed by Guobin {\it et al.}~\cite{guobin2003}, but only in the case of a PCF supporting 10 eigenmodes. The general picture is quite complicated and the degeneracies and order depend on the number of guided eigenmodes~\cite{kuhlmey_mortensen}. In a picture of true eigenmodes this can be understood as follows; for a given number $M$ of guided eigenmodes the eigenvalues are minimized with the constraint that the eigenfields are orthonormal. When the number is changed to, say, $M+1$ the orthogonalization affects all $M+1$ eigenvalues and eigenfields and thus the picture of degeneracies may in general also change compared to the picture for $M$ eigenmodes. 

In our case we numerically find that while ${\boldsymbol E}_3^{(0)}$ and ${\boldsymbol E}_4^{(0)}$ support the full symmetry ${\boldsymbol E}_5^{(0)}$ and ${\boldsymbol E}_6^{(0)}$ do not. Group theory thus predicts that the PCF have a set of modes which are either pair-wise two-fold degenerate ($m=1,2$ and $m=5,6$) or non-degenerate ($m=3$ and $m=4$). This contrasts the situation of a step-index fiber (SIF) in the weakly-guiding approximation where the first higher-order modes are four-fold degenerate. The 3rd and 4th modes are nearly degenerate in the sense that our numerical studies show that the splitting $\beta_4^{(0)}-\beta_3^{(0)}$ is much smaller than the spacings $\beta_5^{(0)}-\beta_4^{(0)}$ and $\beta_3^{(0)}-\beta_2^{(0)}$ to the two neighboring groups of degenerate modes. The grouping of the four higher-order modes into two groups is experimentally reflected in different cut-off wavelengths of the two groups~\cite{folkenberg2003}. Compared to the picture in the weakly-guiding approximation, the splitting of the higher-order modes in PCFs originates from the inversion and six-fold rotational symmetry and the magnitude of the splitting is driven by the high index contrast between air and silica. We note that for a standard fibre, a sufficiently high index contrast will eventually also split the higher-order modes.

\begin{figure}[t!]
\begin{center}
\epsfig{file=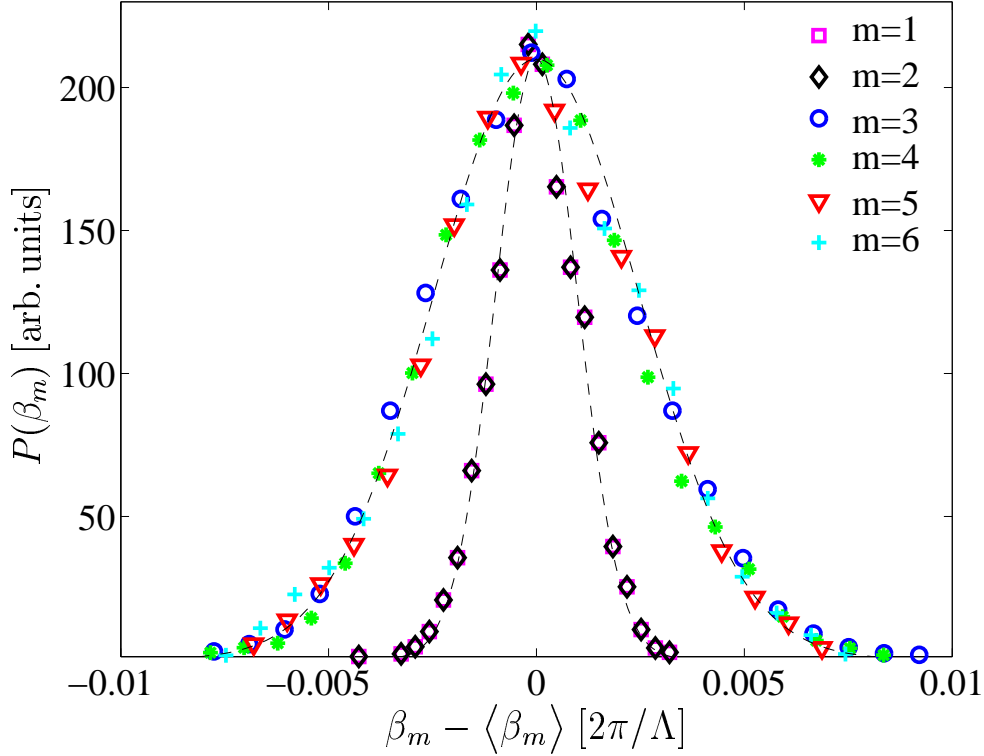, width=1\textwidth,clip}
\end{center}
\caption{Plot of distribution $P(\beta_m)$ for the first six modes $m=1,2,3,4,5$, and $6$. The dashed lines show Gaussian fits.}
\label{fig3}
\end{figure}

From perturbation theory (for the problem of shifting boundaries see Ref.~\cite{johnson2002}) it is obvious that for guided modes the six inner air-holes are of most importance to the optical properties whereas disorder in the air-hole lattice further out in the cladding mainly affects the guided modes in terms of {\it e.g.} leakage loss~\cite{white2001}. Thus, in the search of over-all qualitative effects of disorder it is sufficient to only let the six inner air-holes (see Fig.~\ref{fig1}) have small random displacements $\Delta {\boldsymbol r}_i$ ($i=1,2,\ldots 6$) with respect to the ideal positions of the air-holes as well as small random deviations $\Delta d_i$ in their diameters. We consider the situation where the displacements and deviations in diameter are Gaussianly distributed (zero mean and widths $\sigma_\Lambda$ and $\sigma_d$) and mutually uncorrelated, {\it i.e.}

\begin{eqnarray}
\big<\Delta {\boldsymbol r}_i \big>=\big<\Delta d_i \big>=\big<\Delta{\boldsymbol r}_i \Delta d_j \big>= 0\nonumber\\
 \big<\Delta {\boldsymbol r}_i\Delta {\boldsymbol r}_j \big>=\sigma_\Lambda^2\delta_{ij} \;,\; \big<\Delta d_i\Delta d_j \big>=\sigma_d^2\delta_{ij}.
\end{eqnarray}
In the following we study a large ensemble (more than 1000 members) of fibers numerically.

Since the structural imperfection is typically a small perturbation of the ideal structure we have $\big<\beta_m\big>\simeq \beta_m^{(0)}$. The grouping discussed above thus also holds approximately for the average propagation constants $\big<\beta_m\big> $ as we also have confirmed numerically (not shown). In Fig.~\ref{fig3} we show the distribution  of the propagation constant for the first six eigenmodes at $\lambda=780\,{\rm nm}$ of a PCF with $\Lambda=1.4\,{\rm \mu m}$ and $d/\Lambda=0.67$. For the disorder we have used $\sigma_\Lambda=\sigma_d=0.02\times \Lambda$ corresponding to 2\% structural variations of the six inner air holes. The fundamental mode only has a small fraction of the field near to the air-hole silica interfaces compared to higher-order modes, see row (a) in Fig.~\ref{fig2}. From perturbation theory one thus expects the fundamental mode to be less sensitive to disorder compared to higher-order modes. Indeed, from Fig.~\ref{fig3} it is seen that the distribution $P(\beta_{m})$ has a common width for $m=1,2$ which is significantly smaller than the common width for the higher-order modes ($m=3,4,5,6$).

For a non-zero $\sigma$, symmetries are broken and symmetry-related degeneracies are in principle lifted. For $m=1,2$ this splitting is often referred to as form-birefringence quantified by $\Delta n_{12}=(\beta_1-\beta_2)/k$ where $k$ is the free-space wave number. However, since $\big<{\boldsymbol E}_1^{(0)}\big|\delta\varepsilon \big|{\boldsymbol E}_1^{(0)}\big> \simeq \big<{\boldsymbol E}_2^{(0)}\big|\delta\varepsilon \big|{\boldsymbol E}_2^{(0)}\big>$ for a scalar perturbation $\delta\varepsilon$ there is no significant form-birefringence ($\Delta n_{12}\simeq 0$) to lowest order in perturbation theory. Indeed, for the studied disorder strength we numerically find an extremely narrow distribution. The width is vanishing (within our numerical accuracy) corresponding to a delta-function distribution $P(\beta_1-\beta_2)\simeq \delta(\beta_1-\beta_2)$.

\begin{figure}[t!]
\begin{center}
\epsfig{file=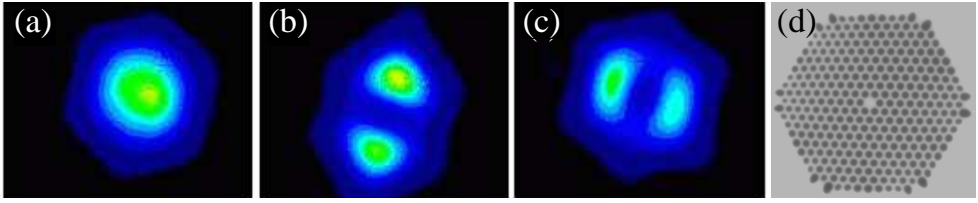, width=1\textwidth,clip}
\end{center}
\caption{Near field images at $\lambda=780\,{\rm nm}$ recorded at the output of 2.0 meters of a PCF. Panels (a)-(c) shows the fundamental and higher-order modes. Panel (d) shows a microscope image of the cross-section of the PCF with $\Lambda\simeq 1.41\,{\rm \mu m}$ and $d/\lambda\simeq 0.66$.}
\label{fig4}
\end{figure}

The effect of disorder on the fields is illustrated in Fig.~\ref{fig2}. Rows (b)-(e) show examples of the electrical field intensity for four random configurations of the inner ring of air-holes. In general, we find that the modes in the ideal structure, row (a), deform into profiles having two maxima with a node in between similarly to the deformation of the modes in elliptical core SIFs~\cite{snyder}. This picture correlates perfectly with the modes observed experimentally in Ref.~\cite{folkenberg2003}. Fig.~\ref{fig4} shows near-fields recorded at $\lambda=780\,{\rm nm}$ for a PCF with $\Lambda\simeq 1.41\,{\rm \mu m}$ and $d/\lambda\simeq 0.66$. Here, the fundamental mode, panel (a), corresponds to the linear combination ${\boldsymbol E}\sim {\boldsymbol E}_1+ {\boldsymbol E}_2$ and for the higher-order modes, panels (b) and (c) corresponds to  ${\boldsymbol E}\sim {\boldsymbol E}_3+ {\boldsymbol E}_4$ and  ${\boldsymbol E} \sim {\boldsymbol E}_5+ {\boldsymbol E}_6$, respectively.

In conclusion we have demonstrated how even weak disorder in the air-hole lattice leads to deformation of the higher-order modes in small-core photonic crystal fibers. In the presence of disorder the higher-order modes resemble those in elliptical core standard fibers. Our findings explain the recently reported higher-order mode profiles and the grouping of higher-order modes correlates with the observation that they have different cut-offs.

\vspace{5mm}

We acknowledge useful discussions with B.~T. Kuhlmey. M.~D. Nielsen and K.~P. Hansen are financially supported by the Danish Academy of Technical Sciences and J. L{\ae}gsgaard by the Danish Technical Research Council.

\vspace{5mm}


\end{document}